\documentclass[11pt]{article}

\usepackage{sint,macros}
\usepackage{amssymb}
\usepackage[dvips]{epsfig}
\font\sixrm=cmr6
\newcommand{\be}{\begin{equation}}
\newcommand{\ee}{\end{equation}}
\newcommand{\bea}{\begin{eqnarray}}
\newcommand{\eea}{\end{eqnarray}}

\begin{document}

\begin{titlepage}

\begin{flushright}
MS-TP-01-16\\
Bicocca-FT-01-25
\end{flushright}

\vskip 1 cm
\begin{center}

{\Large\bf A lattice approach to QCD in the chiral regime}

\end{center}
\vskip 1 cm
\vbox{
\centerline{
\epsfxsize=2.5 true cm
\epsfbox{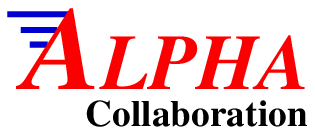}}
}
\vskip 1 cm
\begin{center}
{\large
Michele Della Morte$^{\scriptscriptstyle a}$\footnote{Support by the European
Community's Human Potential Programme under contract RTN1-1999-00246
is acknowledged.},
Roberto Frezzotti$^{\scriptscriptstyle b}$\footnote{I acknowledge
        support and warm hospitality by the Theory Division of CERN
        for the time when the work presented in this contribution was done.} 
and 
Jochen Heitger$^{\scriptscriptstyle c}$
}
\vskip 1.5cm
$^{\scriptstyle a}$
DESY-Zeuthen,\\
Platanenallee 6, D-15738, Zeuthen, Germany
\vskip 2.5ex
$^{\scriptstyle b}$
Universit\`a di Milano-Bicocca, Dipartimento di Fisica,\\
Piazza della Scienza 3, I-20126 Milano, Italy
\vskip 2.5ex
$^{\scriptstyle c}$
Westf\"alische Wilhelms-Universit\"at M\"unster,
Institut f\"ur Theoretische Physik,\\
Wilhelm-Klemm-Strasse 9, D-48149 M\"unster, Germany
\vskip 1.5cm
{\bf Abstract}
\vskip 0.7ex
\end{center}

Non-perturbative lattice studies of QCD in the chiral
thermodynamic regime, where chiral symmetry is spontaneously broken,
require to deal with almost quark zero modes in a theoretically
clean and computationally efficient way. We discuss the basic
features and some realistic tests of a formulation, known as
lattice tmQCD, that fulfills these requirements.

\vskip 1.5ex
\vfill

\begin{center}
November 2001
\end{center}

\eject
\vfill
\eject
\end{titlepage}

%%\author{Michele Della Morte\thanks{Support by the European
%%Community's Human Potential Programme under contract RTN1-1999-00246
%%is acknowledged.}\\
%%        DESY-Zeuthen, Platanenallee 6, D-15738, Zeuthen, Germany}
%%
%%\author{\speaker{Roberto Frezzotti}\thanks{I acknowledge 
%%        support and warm hospitality by the Theory Division of CERN 
%%        for the time when the work presented in this contribution was done.} \\
%%        Univ. of Milano Bicocca, Dip. di Fisica, 
%%        Piazza della Scienza 3, I-20126, Italy                           
%%       }                       
%%
%%\author{ Jochen Heitger \\
%%        WWU M\"unster,
%%                 Institut f\"ur Theoretische Physik,
%%         Wilhelm-Klemm-Str. 9, \\ D-48149 M\"unster, Germany}
%%              
%%
%%\abstract{
%%Non-perturbative lattice studies of QCD in the chiral 
%%thermodynamic regime, where chiral symmetry is spontaneously broken, 
%%require to deal with almost quark zero modes in a theoretically
%%clean and computationally efficient way. We discuss the basic
%%features and some realistic tests of a formulation, known as
%%lattice tmQCD, that fulfills these requirements.
%%}
%%
%%
%%\begin{document}
%%

\section{INTRODUCTION}

Lattice field theory is known to provide a rigorous and systematically 
improvable way of studying QCD in the non-perturbative regime
by means of Monte Carlo simulations. As in any numerical method, the
coexistence of widely separated energy scales poses severe practical
problems: for instance, the energy-momentum resolution $a^{-1}$, which is
induced by the lattice regularization, should be kept much larger than
both the typical hadron mass scale, say $\sim 500$~MeV, and the external
momenta of the correlation functions that are computed. Moreover, the
approximate and spontaneously broken flavour chiral symmetry that is
exhibited by the strong interactions entails the need of performing
at least a few simulations of lattice QCD in large volume and with 
small quark masses\footnote{This task is significantly alleviated
by studying the renormalization problems in finite volume by means of
renormalization group and finite-size scaling techniques, e.g. in 
the Schr\"odinger functional scheme.}: 
%%%\cite{Luscher_LH97}.} :
a fully realistic setup would require a lattice of spatial volume $L^3$, with
$M_\pi L \ge 5$ and $M_\pi \sim 140$~MeV. 

The relevance of the approximate chiral symmetry for the low energy
QCD amplitudes can hardly be overestimated: it determines many aspects
of the dynamics, as shown by the chiral effective Lagrangian approach,
and strongly constrains the operator mixings, which on the lattice
may be particularly severe due to the dimensionful ultraviolet cutoff 
$a^{-1}$. The realization of the chiral symmetry on the lattice is 
known to be delicate since the pioneering work by Wilson \cite{wilson74}
and in most cases the full flavour chiral symmetry is recovered only
in the continuum limit. The highly remarkable exception is represented
by those lattice formulations where the Dirac operator satisfies the 
Ginsparg-Wilson relation \cite{GW_rediscov}, which in turn entails
an exact flavour chiral invariance at finite lattice spacing.
The lattice regularizations with Ginsparg-Wilson quarks certainly
simplify a lot the construction of renormalized operators, especially those
relevant for the weak effective Hamiltonian, at the price however of a big 
overhead in the computational effort. In many cases one can avoid such
an overhead by working within the framework of Wilson quarks, which seems
to be more flexible and powerful than believed till few years ago, as we
try to argue in the following. In general, the lattice regularization 
of the quark sector should be chosen with care depending on the physical 
applications and the related renormalization problems.

In this contribution we focus on a lattice formulation of QCD,
based on Wilson quarks and known as twisted mass QCD (tmQCD), that
is particularly suited for dealing with the $u$ and $d$ quarks.
After discussing in Section~2 how the simulations of lattice QCD account
for the contribution of quark zero modes, we illustrate in Section~3 the 
basic features of lattice tmQCD. In Section~4 we briefly 
report on an exploratory non-perturbative study of
O($a$) improved lattice tmQCD, which reaches a pseudoscalar to vector
meson mass ratio of
$M_{\rm PS}/M_{\rm V} \simeq 0.47(1)$ 
and represents a successful test of  
several aspects of our approach.

\section{Lattice QCD and physical quark zero modes}

We assume that the reader is familiar with the lattice regularization
of QCD introduced by Wilson \cite{wilson74} and for simplicity we consider
the theory with two mass-degenerate quark flavours. 
In this case the lattice action with Wilson quarks reads
\be \label{Wils_std_act}
S = S_{\rm g}[U; g_0^2] + a^4 \sum_{x} \chibar(x) 
\left[ \left( \frac{1}{2} \gamma \cdot (\nabla+\nabla^*)
-a \frac{1}{2} \nabla^* \cdot \nabla +m'_0 \right)
\chi\right](x) \; ,
\ee 
where $S_{\rm g}$ is the pure gauge action, $\chi$ denotes the 
doublet of quark fields and $ v \cdot w \equiv v_\mu w_\mu$, while
$\nabla_\mu = \nabla_\mu[U]$ and $\nabla^*_\mu=\nabla_\mu^*[U]$ stand for
the forward and backward covariant derivatives on the lattice. 
The hard breaking of the flavour non-singlet axial generators induced by the
Wilson term can be compensated up to O($a$) effects by tuning the parameter
$m'_0$ and the coefficients that parameterize any chirality-violating operator 
mixings \cite{bmmrt}. The leading cutoff effects can be removed
via the on-shell O($a$) improvement \cite{O(a)_impr} \`a la Symanzik. 

%The hard breaking of the flavour non-singlet axial generators induced by the
%Wilson term can be compensated up to O($a$) effects by tuning the parameter
%$m'_0$ and the coefficients that parameterize any chirality-violating operator 
%mixings \cite{bmmrt}. 
%In particular, the massless theory is obtained 
%for $m'_0 = \mc(g_0^2)$. Lattice regularizations with Wilson quarks are 
%widely used for non-perturbative studies of QCD, owing 
%to the clean theoretical status, the absence of complications in building
%interpolating fields with given quantum numbers, the simple next-neighbours 
%form of the action and the possibility of removing the leading cutoff effects
%via O($a$) improvement \`a la Symanzik \cite{O(a)_impr}.  

\subsection{Valence and sea quarks}

In any correlation function to be evaluated via Monte Carlo simulations,
due to the huge dimensionality of the vector space spanned by the Grassmann
variables in the Euclidean path integral, it is customary to integrate
out analytically the fermionic degrees of freedom. In the case of the 
correlator of two local pseudoscalar densities with isospin index $a=1$,  
\be \label{pion_corr1}
C_{\rm PS}^{11}(x-y) \equiv
- \langle P^1(x) P^1(y) \rangle = - {\cal Z}^{-1} \int {\rm d}U {\rm d}\chi 
{\rm d}\chibar \exp(-S) \; [\chibar \gamma_5 \frac{\tau^1}{2} \chi](x)
[\chibar \gamma_5 \frac{\tau^1}{2} \chi](y) \; ,
\ee 
this procedure yields
\bea \label{pion_corr2}
 C_{\rm PS}^{11}(x-y) & = & 
 \int {\rm d}U \; P[U;g_0^2,\mq'] \; \cdot \nonumber \\
 & \cdot & {\rm tr}\left\{ [D_{\rm W,c} + \mq']^{-1}(x,y)
\gamma_5 \frac{\tau^1}{2} [D_{\rm W,c} + \mq']^{-1}(y,x)
\gamma_5 \frac{\tau^1}{2} \right\} \; ,  \nonumber \\
P[U;g_0^2,\mq'] & = &  {\cal Z}^{-1} \exp(-S_{\rm g}[U,g_0^2])
\det( \{D_{\rm W,c} + \mq'\}[U] ) \; \geq \; 0 \; ,
\quad\quad\quad\quad\quad
\eea
where\footnote{The symbol ${\rm tr \{ \dots \} }$ denotes 
the trace over flavour, colour and spin indices, whereas the 
symbol $\det \{ \dots \} $ stands for the determinant with respect
to all indices, including the space-time ones.} ${\cal Z}$ 
denotes the Euclidean partition function and
$D_{\rm W,c} = D_{\rm W,c}[U]$ is the critical (two-flavour) Wilson-Dirac 
operator:
\be \label{Wc}
D_{\rm W,c} = \frac{1}{2} \gamma \cdot (\nabla+\nabla^*)
- \frac{a}{2} \nabla^* \cdot \nabla + \mc \; .  
\ee
The parameter $\mq' = m'_0 - \mc$ is hence proportional to the
renormalized quark mass that appears in the PCAC Ward identity:
the massless theory is obtained
for $m'_0 = \mc(g_0^2)$.
Gauge configurations $U$ are generated via suitable algorithms
with probability $P[U;g_0^2,\mq']$, and on each configuration $U$
the quark propagator from the lattice site $x$ to the site $y$,
$[D_{\rm W,c} + \mq']^{-1}(y,x)$, can be computed --for
fixed $x$-- by solving a linear system.
By expanding the correlator~(\ref{pion_corr1}) around the trivial 
perturbative vacuum, one can easily check that in eq.~(\ref{pion_corr2}) 
the trace term involving quark propagators corresponds to valence
quark diagrams dressed with any kind of purely gluonic corrections, whereas 
the term 
$\det( D_{\rm W,c} + \mq' )$ accounts for the sea quark corrections to
the aforementioned dressed valence quark diagrams. 

\subsection{Quenched and partially quenched lattice QCD}

It is well-known that with the established simulation techniques
the simulation of the full theory, including the sea quark effects, 
has a very high computational cost, which quickly increases as
the pion mass is decreased towards realistic values.
On the other hand, it is technically trivial to choose different values 
for the parameter $\mq'$ that appears in the fermionic determinant 
and its counterpart in the inverse Dirac operator: 
$ m'_{\rm q, sea} \neq m'_{\rm q, val}$. A moment of thought reveals
that such a modification of lattice QCD corresponds to a statistical 
model with extra spin-$1/2$ ghost fields, which
is local and renormalizable by power counting
but violates reflection positivity (see e.g. Ref.~\cite{BeGo}).  
The general case $ m'_{\rm q, val} 
\neq m'_{\rm q, sea}$ is referred to as partially quenched QCD, whereas
the particular case $ |m'_{\rm q, val}| < |m'_{\rm q, sea}| = \infty$
corresponds to the well-known quenched approximation. 

It should be noted that these approximations in general break down as the
valence quark mass $m'_{\rm q, val} \to 0$ in the thermodynamic limit.
In this regime the flavour chiral symmetry is spontaneously broken:
if the chiral condensate is to be non-vanishing, 
the gauge configurations carrying zero modes of the Dirac operator
must receive a finite weight in the Euclidean path integral, even
in the limit (taken at infinite spatial volume) 
$ m'_{\rm q, sea} = m'_{\rm q, val} \to 0$ \cite{CaBa}. It is clear that
in the full theory the integration over the fermionic variables on
any gauge background can not yield divergences: in presence of
quark zero modes the infinities in the quark propagators must be
compensated by the zeros in the fermion determinant, see 
eq.~(\ref{pion_corr2}). Such a delicate compensation is no
longer guaranteed if the condition 
$ m'_{\rm q, sea} = m'_{\rm q, val}$
is violated: fermionic observables may hence diverge on gauge
configurations carrying quark zero modes, which causes the
breakdown of the quenched and partially quenched approximations.
 
However, when working sufficiently away from the thermodynamic chiral limit, 
the quenched --or partially quenched-- approximation is expected to be 
reasonably accurate, at least for those quantities that are not very
sensitive to sea quark effects. An example is given by the ratios of hadron 
masses \cite{CPPACS_hadspe}, with the $\eta'$-meson mass being the most striking 
exception. Indeed, given the high computational cost of simulating
the full theory, the quenched approximation has been widely used
as a testbed for lattice techniques and for first non-perturbative 
estimates of quantities such as renormalized couplings, hadron masses 
and matrix elements, order parameters of phase transitions.
The chiral effective Lagrangian for quenched QCD has also been worked 
out \cite{BeGo} with the aim of identifying and parameterising the  
deviations from the full theory close to the chiral limit.

Analogous remarks apply for the partially quenched approximation,
the quality of which depends on the ratio $ m'_{\rm q, sea}/m'_{\rm q, val}$. 
In particular, it has been remarked that the low-energy (Gasser-Leutwyler)
constants of partially quenched QCD with
$N_{\rm f}$ quark flavours coincide with those of the fully unquenched 
theory, provided that all the quark flavours are light enough for the
chiral perturbation theory to be applicable \cite{ShaSho}. 
In the physically relevant
cases, which are $N_{\rm f}=3$ and --to some extent-- $N_{\rm f}=2$, 
varying $m'_{\rm q, val}$ while keeping fixed 
$ m'_{\rm q, sea}$ can then be very convenient, since it allows
to investigate the dependence of the observables on $m'_{\rm q, val}$
without performing many unquenched simulations.

\section{Lattice tmQCD}

Lattice tmQCD is an extension
of the widely used formulation with Wilson quarks,
from which it differs in that the physically non-vanishing quark mass term
is in general not aligned with the Wilson term in the flavour chiral space.
This simple modification brings definite advantages concerning
the simulations with light quarks --especially in the quenched or
partially quenched case-- and the renormalization properties of some
phenomenologically important quantities, such as the leptonic decay
amplitude of pseudoscalar mesons or the mixing amplitude in the
$K^0$-$\overline{K}^0$ system \cite{paper1,FS01,BK01}.

\subsection{Lattice Wilson quarks and spurious zero modes}

While the considerations of Subsections~2.1--2.2, 
apply to any sensible lattice regularization of QCD,
working with Wilson fermions entails a further  
technical problem that renders particularly difficult 
--or even impossible-- the simulations with light quarks.
The problem arises whenever the quenched
(or partially quenched) sample of configurations,
as determined by a given choice of the bare parameters, includes gauge
backgrounds on which the critical Wilson-Dirac operator $D_{\rm W,c}$,
eq.~(\ref{Wc}), has one or more eigenvalues with {\em negative} 
real part ${\rm Re}(\lambda) <0$
and (almost) {\em zero} imaginary part. Under these conditions, the
massive Dirac operator $D_{\rm W,c} + m'_{\rm q,val}$ is singular for values
of $m'_{\rm q,val}>0$ as soon as $m'_{\rm q,val} + {\rm Re}(\lambda) =0$ 
for some of the real negative eigenvalues $\lambda$. 

The fermionic observables, which involve
$[D_{\rm W,c} + m'_{\rm q,val}]^{-1}$, may receive (almost) divergent 
contributions on the aforementioned gauge backgrounds, spoiling
the expected decrease of the statistical errors with the number
of independent measurements \cite{Bardeen_et_al}. 
As an example of this phenomenon, we show
in Fig.~\ref{fig_fP24} 
the Monte Carlo history of the normalized relative standard deviation
for the pion channel correlator $f \equiv
\fp^{11}(x_0)$ at $x_0=24a=T/2$, see Section 4.1. The normalization
of the standard deviation is such that it should approach 
a constant in the limit of infinite statistics: in the case of
the standard Wilson regularization, which corresponds to the open
symbols in Fig.~\ref{fig_fP24}, the problem is apparent.
Via the combined effect of lattice artifacts and statistical fluctuations,
"spurious" quark zero modes\footnote{These spurious quark zero modes 
should not be confused with 
the quark zero modes that play an important physical role in the
thermodynamic chiral limit of renormalized QCD.}
anticipate at non-vanishing values of $\mq'$ the breakdown of the 
quenched or partially quenched approximation.
%Via a conspiration of lattice artifacts induced
%by the Wilson term and statistical fluctuations, 
%the breakdown of the quenched or partially quenched approximation
%is anticipated at non-vanishing values of $\mq'$. 
%This phenomenon is in contrast with the behaviour that one would
%expect in the continuum theory where a finite quark mass provides an
%infrared cutoff for fermionic observables, and the associated quark zero
%modes will hence be called "spurious". They should not be confused with
%those quark zero modes that play an important physical role in the
%thermodynamic chiral limit of renormalized QCD.  

In practice, in the quenched case 
the rate of occurrence of gauge backgrounds with spurious 
quark zero modes  --also called "exceptional configurations"-- depends on
the values of $m'_{\rm q,val}$ and $g_0^2$, on the physical linear size 
$L$ of the lattice and on various details of the lattice regularization. 
The rate tends to increase when decreasing $m'_{\rm q,val}$ and
increasing $g_0^2$ and $L/a$, as well as when switching on the
coefficient, $\csw(g_0^2)$, of the 
counterterm that is needed for the on-shell O($a$) improvement of the
fermionic action. For instance, in the regularization with plaquette
gauge action and non-perturbatively O($a$) improved Wilson quarks, the
problem is felt for $L \geq 1.5$~fm at values of $g_0^2$ corresponding
to $a \sim 0.1$~fm and at valence quark masses that are about half the 
strange quark mass. A similar problem is also expected in partially 
quenched simulations, although with a lower rate depending
on the ratio $m'_{\rm q,val}/m'_{\rm q,sea}$, and has indeed been
observed, see e.g. Ref.~\cite{pq_exceptionals}.

In the fully unquenched case, one can expect troubles at the algorithmic
level with light Wilson quarks on rather coarse lattices. This is because
the state-of-the-art algorithms implement stochastically the fermion
determinant that appears in the probability measure for the gauge 
configurations: almost exceptional configurations may hence be proposed, but
are then almost certainly rejected. 
In this process however the simulation algorithm, 
e.g. the standard HMC one\footnote{Evidence for a large increase of the
fermionic force at small quark mass values is reported in 
Ref.~\cite{unq_exceptionals}.}, undergoes a severe slowing-down,
due to a decrease in the acceptance rate and an increment of the condition
number of the Dirac matrix before the accept/reject test.

\subsection{Action and symmetries} 

As already known since 1989 \cite{AoGo}, the problem with spurious
quark zero modes is absent in the two-flavour theory if one considers 
the action   
\be \label{Wils_tm_act1}
  S_{\rm W}[U,\psibar,\psi] = S_{\rm g} + a^4\sum_{x}
  \psibar (x) [(D_{\rm W, c} +\mq +i\muq\gamma_5\tau^3 ) \psi] (x) \; , 
\ee
where $\psi$ is a flavour quark doublet, the matrix $\tau^3$
acts in the flavour space and the boundary conditions for
finite-volume systems may remain unspecified for a while.
Since $\mc$ is known in practice with finite precision, the 
exactly known bare parameters are $g_0^2$, $\muq$ and $m_0 = \mc + \mq$. 
It is easy to see that, as long as $\muq \neq 0$, 
on any gauge background the lowest eigenvalue 
of the Hermitean square of the matrix 
$(D_{\rm W, c} +\mq +i\muq\gamma_5\tau^3)$ 
is bounded from below by $(a\muq)^2$.

The action~(\ref{Wils_tm_act1}) represents a 
sensible regularization of QCD with
$N_{\rm f}=2$ mass-degenerate quark flavours, but in a quark field basis 
that is chirally twisted with respect to the standard one \cite{paper1}. 
To illustrate this point, let us focus on the simple case $\mq=0$, $\muq 
\neq 0$, and consider the very {\em same} lattice theory in the
standard quark field basis, which is obtained by a suitable axial rotation
with generator $\tau^3$:  
\be \label{change_latbas}
\chi = \exp\left(i\omega\gamma_5 {\tau^3 \over 2}\right) \psi
\; , \quad\quad\quad
\chibar = \psibar \exp\left(i\omega\gamma_5 {\tau^3 \over 2}\right)
\; , \quad\quad\quad
 \omega = {\rm arctan}{\muq \over \mq} = {\pi \over 2} \; .
\ee
The action~(\ref{Wils_tm_act1}) then reads
\be \label{Wils_tm_act2}
  S_{\rm W}[U,\chibar,\chi] = S_{\rm g} + a^4\sum_{x}
  \chibar (x) [(D_{\rm tmW, c} + \muq ) \chi] (x) \; , 
\ee
where
\be \label{D_tmW_M}
D_{\rm tmW, c} = \frac{1}{2} \gamma \cdot (\nabla+\nabla^*)
+i \frac{a}{2} \gamma_5 \tau^3 \nabla^* \cdot \nabla 
- i\gamma_5 \tau^3 \mc \; .
\ee
The connection, eq.~(\ref{change_latbas}), between the two lattice
quark bases makes obvious that the chiral limit is obtained for
$\mq = \muq =0$ with $\mc$ being the usual function of $g_0^2$.  
Inspection of the symmetries of the action~(\ref{Wils_tm_act2})
shows that in the chiral limit one vector and
two axial generators --out of the six generators of the flavour
chiral group-- are preserved by the lattice regularization,
while parity is preserved only up to a flavour exchange ($P_{\rm F}$ 
symmetry) and all other symmetries are as usual with Wilson fermions. It 
follows that $\muq$ is {\em multiplicatively} renormalized,
$\mu_\rmR = Z_\mu(g_0^2) \, \muq$, while 
$m_\rmR = \zm (g_0^2) \, \mq$.
Power counting renormalizability and the recovery
of the full flavour chiral symmetry in the continuum limit \cite{bmmrt}, 
together with the exact $P_{\rm F}$ invariance, imply that parity must
also be recovered in the continuum limit. In contrast with other approaches
based on the action~(\ref{Wils_tm_act1}) \cite{Bardeen_et_al}, the correlation
functions of the massive QCD are obtained at finite $\muq$, so that the 
problem with the spurious quark zero modes is certainly solved.

In the general case, where both $\mq$ and $\muq$ are non-vanishing,
the situation is fully analogous, but the relation between the two 
lattice quark bases involves an angle $\omega \neq \pi/2$. However,
owing to the complications arising from the Wilson term, it is 
convenient to perform first the renormalization (and possibly the O($a$) 
improvement) of the correlation functions in a given quark basis and then 
transform to the "physical" basis, i.e. the one where the quark mass is coupled
to the singlet scalar density. The first step implies that the continuum
limit is approached at fixed values of $g^2_\rmR$, $\mr$, $\mur$ and
the normalization conditions for the composite fields. Concerning
the second step, for a wide class of renormalization schemes, the relation 
between the renormalized fields in two different bases that are related
via a non-singlet axial rotation takes the same form as at the classical 
level \cite{paper1}.
In terms of polar quark mass coordinates, 
\be \label{polar_mass}
\tan \alpha = { \mur \over \mr} \, , \quad
M_\rmR= \sqrt{\mur^2 + \mr^2} \, ,
\ee
the angle $\alpha$ identifies the quark basis and hence
specifies the axial rotation with generator $\tau^3/2$
that allows to combine the renormalized correlation functions so
to obtain finite correlators with given 
continuum quantum numbers\footnote{This point of view might lead to a 
reinterpretation of the "spontaneous breakdown of parity and isospin" 
in lattice QCD with Wilson fermions \cite{Aoki_phase}, which was 
observed by employing the action~(\ref{Wils_tm_act1}).}.
On the latter correlators the partial breaking of isospin and parity   
is an effect of order $a\mu_\rmR$ and represents no serious drawback.
    
If one identifies the two quark flavours with the $u$ and $d$ quark,
the tiny mass difference between them can   
safely be taken into account by means of chiral perturbation theory.
Alternatively, lattice tmQCD can also be formulated for a doublet of 
non-degenerate quarks, whilst retaining the protection against
spurious quark zero modes~\cite{RFnotes}.
Heavier flavours of Wilson quarks can be added e.g. in the usual way
to the lattice tmQCD action for $N_{\rm f}=2$.

\section{A test-study of lattice tmQCD with light quarks and in large volume}

We present here some preliminary results of a high-statistics
exploratory study of lattice tmQCD in the thermodynamic chiral
regime. The study aimed at testing the absence of exceptional
configurations, the computational cost and the magnitude of 
cutoff effects for a few typical observables. We adopt in the following
the notation
of Ref.~\cite{paper3}, to the equations of which we refer  
with the prefix "I", and postpone
many technical details to a forthcoming publication \cite{paper4}.

\subsection{Observables and simulations}
In this test-study we choose to work in the quark basis of
action~(\ref{Wils_tm_act1}) and 
implement the {\em non-perturbative} O($a$) improvement of the action and the
relevant operators along the lines of Ref.~\cite{paper2}. 
Attention is restricted to the pseudoscalar and vector meson
masses, $M_{\rm PS}$ and $M_{\rm V}$, the pseudoscalar (leptonic) decay constant,
$F_{\rm PS}$, and the polar quark masses, $M_\rmR$ and $\alpha$, which
are defined according to eqs.~(I.3.3)--(I.3.20) for $\mur$ and $\mr$
and eq.~(\ref{polar_mass}). 

%%%%%%%%%%%%%%%%%%%%%%%%%%%%%%%%%%%%%%%%%%%%%%%%%%%%%%%%%%%%%%%%%%%%%%%
%%\TABLE{ 
\begin{table}[htb]
\vspace{0.25cm}
\begin{center}
\begin{tabular}{cccccccc}
\hline
\small
$\!\!\!\!$Set, $\beta$  & $\!\!L/a$, $T/L$ & $\!\!N_{\rm meas}$ & $m_\rmR/M_\rmR$ & $M_{\rm R} r_0$ & $M_{\rm PS}r_0$ & $F_{\rm PS}r_0$ & $M_{\rm V}r_0$ \\
\hline
$\!\!\!\!$A1, 6.0  & 16, 2 &  $\!\!$650 & $-$0.016(3) & 0.2729(15)  & 1.711(7) & 0.455(5) & 2.662(40)    \\
$\!\!\!\!$A1',6.0 & 16, 3 &  $\!\!$650 & $-$0.016(3) & 0.2729(15)  & 1.714(6) & 0.455(6) & 2.656(42)    \\
$\!\!\!\!$A2, 6.2  & 24, 2 &  $\!\!$535 & $-$0.014(2) & 0.2558(16)  & 1.623(8) & 0.456(5) & 2.557(32)    \\
$\!\!\!\!$B1L, 6.0  & 24, 2 &  $\!\!$260 &  0.017(3) & 0.1949(11)  & 1.452(6) & 0.432(6) & 2.517(35)    \\
$\!\!\!\!$B1, 6.0   & 16, 2 &  $\!\!$535 &  0.001(3) & 0.1949(11)  & 1.455(8) & 0.428(5) & 2.513(47)    \\
$\!\!\!\!$B2, 6.2   & 24, 2 &  $\!\!$300 & $-$0.004(4) & 0.1962(12)  & 1.420(9) & 0.436(7) & 2.462(41)    \\
$\!\!\!\!$C, 6.0    & 24, 2 &  $\!\!$260 &  0.083(5) & 0.1205(7)   & 1.160(6) & 0.401(6) & 2.485(59)    \\

\hline
\end{tabular}
\caption{Statistics and renormalized quantities obtained in our simulations,
which are identified by a label and the value of $\beta=6/g_0^2$. 
The statistics is specified by the number of measurements,
$N_{\rm meas}$, on almost independent gauge configurations, while the values
of $m_0$ and $\muq$ can be found in Ref.~[20].}
%%%\cite{paper4}.} 
\label{tab1}
\end{center}
\end{table}
%
%%%%%%%%%%%%%%%%%%%%%%%%%%%%%%%%%%%%%%%%%%%%%%%%%%%%%%%%%%%%%%%%%%%%%%%

We work in the {\em quenched} approximation considering systems of physical
size $L^3 T$ with $L$ such that $M_{\rm PS}L \geq 4.5$ to suppress 
finite volume effects. In practice we take $L=1.5$~fm or $L=2.2$~fm and 
$T/L=2 \div 3$.  
Following Ref.~\cite{paper3}, we impose boundary conditions of
Schr\"odinger functional \makebox{(SF)} type and compute the SF correlators
\be 
\fa^{11}(x_0)\, , \quad \fp^{11}(x_0)\, , \quad
\fv^{12}(x_0)\, , \quad 
\kv^{11}(x_0)\, , \quad \kt^{11}(x_0)\, ,\quad
\ka^{12}(x_0)\, , \quad f_1^{11} \; .
\ee
We hence construct the linear combinations of renormalized and O($a$) improved 
SF correlators that correspond to operator insertions (at time $x_0$) with
well defined continuum quantum numbers, see eqs.~(I.3.10)--(I.3.11)
with $\alpha$ given by eq.~(\ref{polar_mass}). Namely, the correlators 
$[\fap^{11}]_{\hbox{\sixrm R}}(x_0)$ and $[\fpp^{11}]_{\hbox{\sixrm R}}(x_0)$
correspond to the insertion of the isotriplet pseudoscalar operators 
$(A'_{\rmR})^1_0$ and $(P'_{\rmR})^1$, while the correlators 
$[\kvp^{11}]_{\hbox{\sixrm R}}(x_0)$ and $[\ktp^{11}]_{\hbox{\sixrm R}}(x_0)$ 
correspond to the insertion of the isotriplet vector operators
$(V'_{\rmR})^1_k$ and $(T'_{\rmR})^1_{k0}$.
In the limit of large $x_0$ and large $(T-x_0)$ and up to cutoff effects, 
these non-vanishing correlators are expected to be dominated by the lowest
isotriplet pseudoscalar and vector meson states.

%%\EPSFIGURE
\begin{figure}[htb]
\begin{center}
\epsfig{file=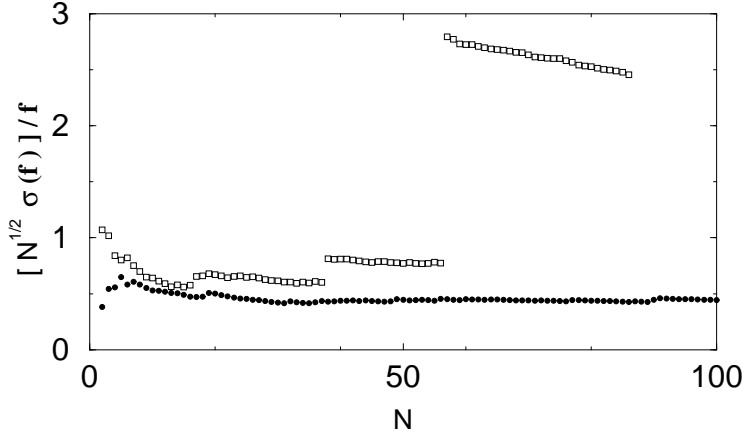,width=6.0cm,angle=-90}
\caption{
The square root of the relative a priori variance of $f=\fp^{11}(x_0=24a)$
versus the number of measurements $N$: data from the simulation C
(filled symbols) and another simulation with the same values of
$\beta$ and $M_\rmR$, but $\alpha=0$ (open symbols).
        }   
\label{fig_fP24}
\end{center}
\end{figure}
%%}

An overview of our simulation parameters, statistics and preliminary results 
for renormalized quantities is given in Table~1. The renormalized gauge 
coupling $g^2_\rmR$ is eliminated in favour of the length scale $r_0$
\cite{sommer_r_0}, which is known to be about 0.5~fm, while the lattice 
spacing value corresponding to $\beta= 6 \; (6.2)$ is 
$a \sim 0.093 \; (0.068)$~fm. 
Our most critical simulation (set C, $\beta=6$),
where we employed a CGNE solver
for the SSOR-preconditioned version of the
Dirac matrix $(D_{\rm W, c} +\mq +i\muq\gamma_5\tau^3 )$, required
$\sim 230$~GFlops~$\times$~day.

%%\FIGURE{
\begin{figure}[htb]
\begin{center}
\hspace{-1.5cm}\epsfig{file=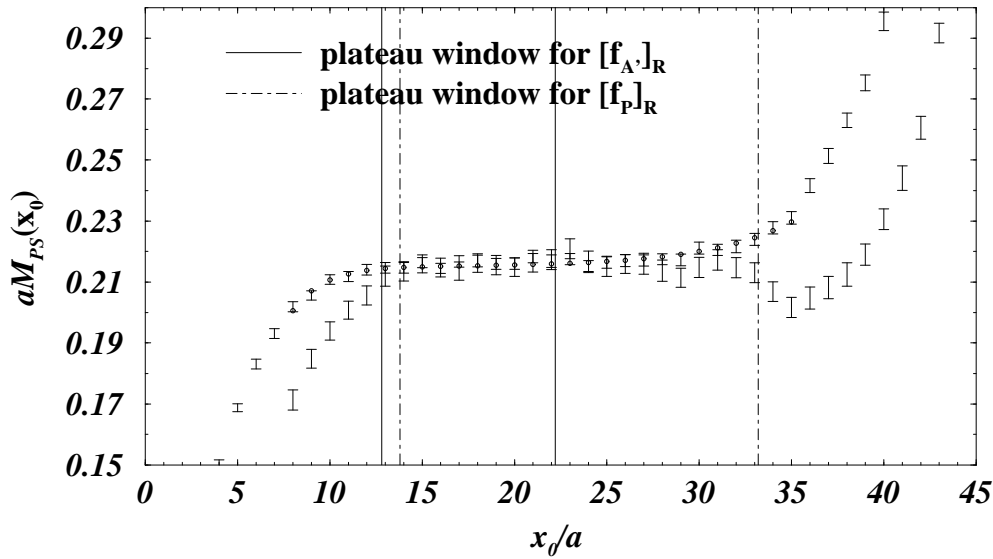,width=7.5cm,angle=-90}
\caption{
Pseudoscalar effective masses extracted from the correlators
$[\fp]_\rmR \equiv [\fpp^{11}]_\rmR$ and 
$[\fap]_\rmR \equiv [\fap^{11}]_\rmR$ for the simulation C.
The tiny circles denote our (good) fit to
the effective masses from $[\fap]_\rmR$:
at large $x_0$ a peculiar O($a\muq$) contribution is present.
        }
\label{fig_meff}
\end{center}
\end{figure}
%%}

%\DOUBLEFIGURE
%{fp24_conv.eps,width=4.0cm,angle=-90} 
%{Fig5.eps,width=3.5cm,angle=-90}
%{
%The square root of the relative a priori variance of $f=\fp^{11}(x_0=24a)$ 
%versus the number of measurements $N$: data from the simulation C 
%(filled symbols) and another simulation with the same values of 
%$\beta$ and $M_\rmR$, but $\alpha=0$ (open symbols).
%\label{fig_fP24} 
%}
%{
%Pseudoscalar effective masses extracted from the correlators
%$[\fpp^{11}]_\rmR$ and $[\fap^{11}]_\rmR$ for the simulation C.
%The tiny black circles denote our (good) fit to 
%the effective masses from $[\fap^{11}]_\rmR$:
%at large $x_0$ a peculiar O($a\muq$) contribution is present.
%\label{fig_meff}
%}

\subsection{Results}

We find that lattice tmQCD
allows, as expected, to safely work in a region
of parameters which would be inaccessible
with ordinary Wilson quarks: see e.g. Fig.~\ref{fig_fP24} as
well as the findings of Ref.~\cite{UK_tmQCD}.
For a given number of independent measurements, the statistical errors on
$M_{\rm PS}$ and $M_{\rm V}$ are comparable, up to a factor of one to three,
to those
found e.g. with domain wall quarks \cite{CPPACS_B_K}. The CPU time
effort, e.g. for the data sets A1, A1' and A2, is in line with
the computational cost for ordinary Wilson quarks.

%\DOUBLEFIGURE
%{Mpi_pro.eps,width=4.2cm,angle=-90}
%{Fpi_pro.eps,width=4.05cm,angle=-90}
%{
%$( M_{\rm PS} r_0 )^2$  versus $M_\rmR r_0$ from
%\mbox{tmQCD} and reanalysis of the data of 
%\mbox{Ref.~\cite{garden}}, including a
%continuum extrapolation (c. l.). 
%%\label{fig_chir1}
%}
%{ 
%The analogous of Fig.~3 for $F_{\rm PS} r_0$.
%Symbols are the same as specified in the legenda of Fig.~3.
%%\label{fig_chir2}
%}

%%\FIGURE{
\begin{figure}[htb]
\begin{center}
\epsfig{file=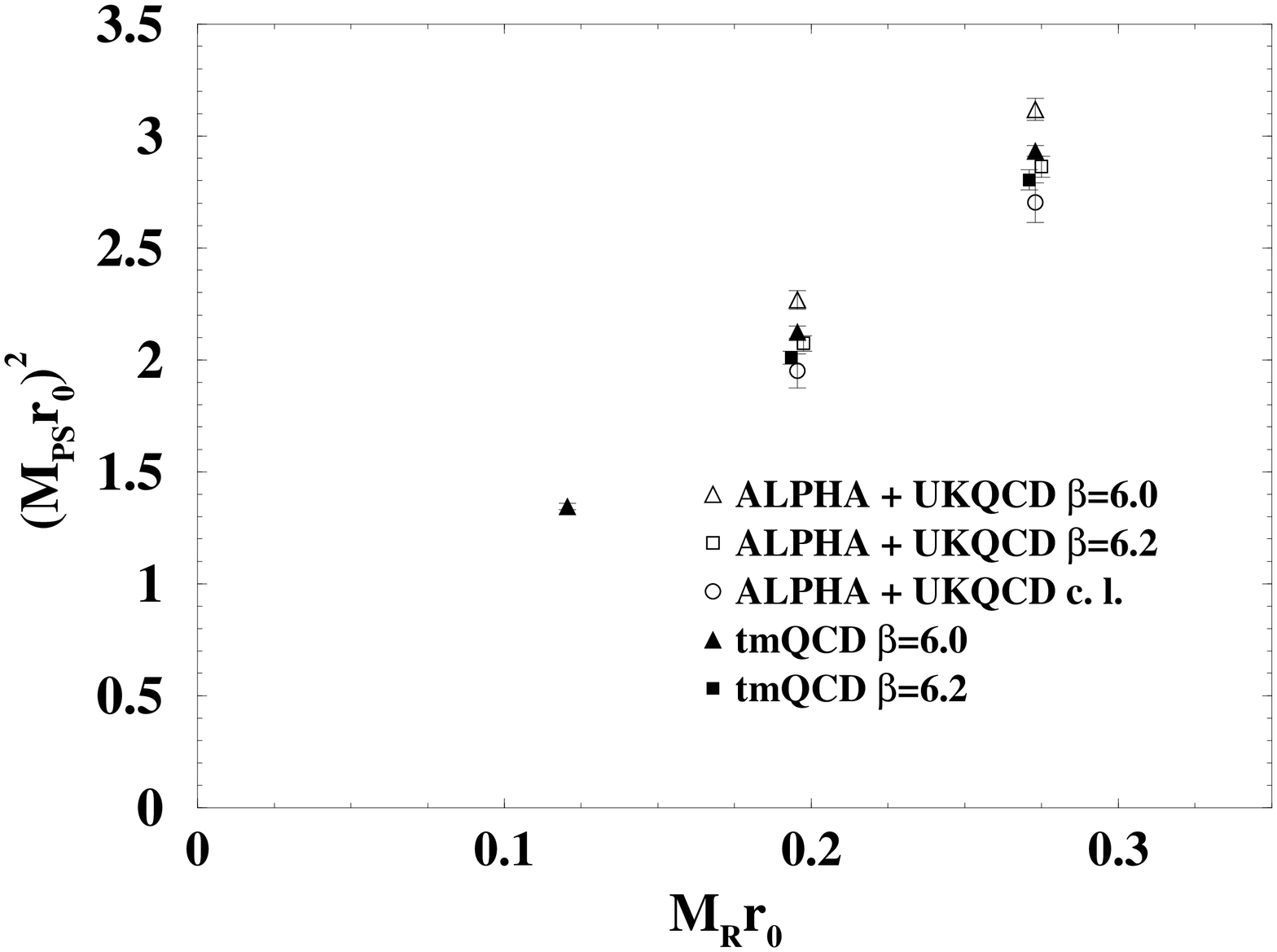,width=7.8cm,angle=-90}
\caption{ 
$( M_{\rm PS} r_0 )^2$  versus $M_\rmR r_0$ from
\mbox{tmQCD} and reanalysis of the data of
%%%\mbox{Ref.~\cite{garden}}, including a
\mbox{Ref.~[23]}, including a
continuum extrapolation (c. l.).
        }
\label{fig_chir1}
\end{center}
\end{figure}
%%}

The partial breaking of parity and isospin that is peculiar of 
lattice tmQCD is found to be a minor problem within our small
statistical errors. In this respect it should be noted that we work 
at small values of $a\muq$, namely
$0.0266 \geq a\muq \geq 0.0117$, and consider observables in
physical channels where the lowest state is lighter than the lowest
state of the corresponding channels with flipped parity and isospin.
While deferring the details of our analysis to Ref.~\cite{paper4}, we
show in Fig.~\ref{fig_meff} an example of effective masses extracted
from SF correlators, where the correlator $[\fap^{11}]_\rmR(x_0)$
receives contributions of order $a\muq$ that are visible at large $x_0$. 
Analogous effects are expected and found to be negligible within
statistical errors for both $[\fpp^{11}]_\rmR(x_0)$ and the 
vector channel correlators.
 
As detailed in Refs.~\cite{paper2,FS01}, we expect the relations among
our observables and the renormalized parameters 
$r_0$ and $M_\rmR$ to be O($a$) improved.
In particular, when working at $\alpha = \pi/2 + {\rm O}(a)$,
which is the case of our study, the knowledge of 
a few counterterms (those with coefficients $\zg$, 
$\mc$ and $\csw$) suffices to obtain 
an O($a$) improved estimate of $F_{\rm PS}$. 
In order to check for the residual scaling violations, we produced data
at $\beta=6.2$ (sets A2 and B2), while keeping $\alpha$, $M_\rmR$ 
and $r_0$ fixed. The small mismatch in $M_\rmR r_0$ for the set A2
was corrected by employing estimates of the dependence
of our observables on $M_\rmR r_0$.

%%\FIGURE{
\begin{figure}[htb]
\begin{center}
\epsfig{file=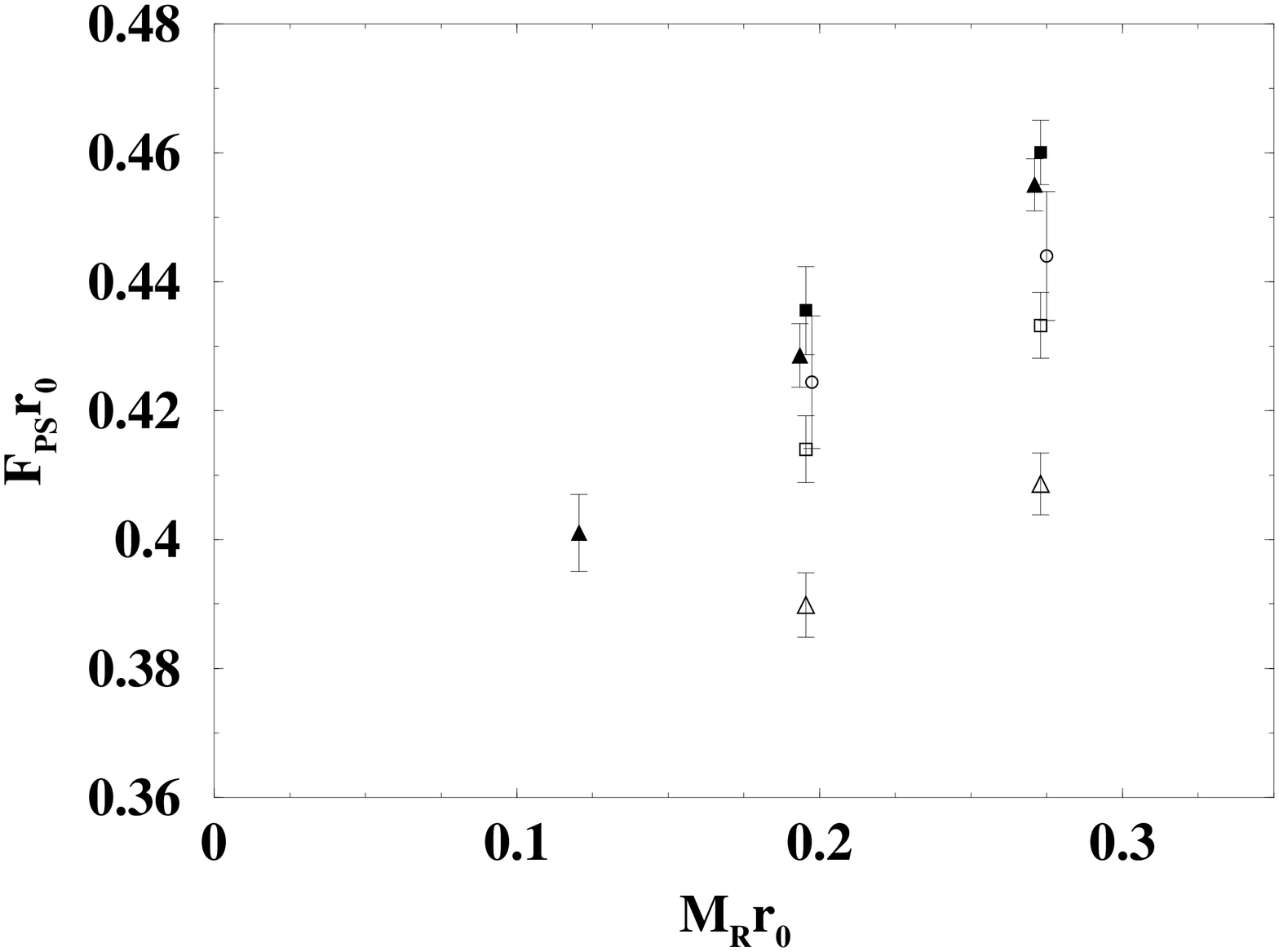,width=7.6cm,angle=-90}
\caption{ 
The analogous of Fig.~3 for $F_{\rm PS} r_0$.
Symbols are the same as in the legenda of Fig.~3.
        }
\end{center}
\label{fig_chir2}
\end{figure}
%%}

We also reanalysed the data of Ref.~\cite{garden}, which were
produced at $\beta=6,6.1,6.2,6.45$, by imposing precisely the same 
renormalization conditions as in this study of tmQCD. We then
performed a continuum extrapolation of these data, assuming a purely quadratic
dependence on $(a/r_0)^2$ and discarding the data at $\beta=6$.
However, the resulting estimate of $F_{\rm PS}$ is not fully O($a$) improved,
as for one of the necessary improvement coefficients, $\ba(g_0^2)$, only the
one-loop estimate could  be used.
The outcome of this exercise is compared with the results from tmQCD
in Figs.~3--4, 
%%%in Figs.~\ref{fig_chir1}--\ref{fig_chir2}, 
omitting the case of $M_{\rm V} r_0$ where
cutoff effects are hardly visible within statistical errors. The
estimators of $M_{\rm PS}$ and $F_{\rm PS}$ that are obtained from 
lattice tmQCD show rather small cutoff effects, which agrees
with the findings of a scaling test \cite{paper3} in intermediate
volume ($L=0.75$~fm).

%%%%%%%%%%%%%%%%%%%%%%%%%%%%%%%%%%%%%%%%%%%%%%%%%%%%%%%%%%%%%%%%%%%%

\section{Conclusions}

Lattice tmQCD is well suited to perform non-perturbative studies of QCD
in the chiral (thermodynamic or finite-volume) regime for all
cases where it is technically sufficient to recover chiral
symmetry in the continuum limit. The framework has been successfully
tested in the quenched approximation and can straightforwardly be
extended beyond it.

The tmQCD project is part of the ALPHA
Collaboration research programme. We acknowledge the very pleasant
collaboration with P.A.~Grassi, S.~Sint and P.~Weisz, as well as
fruitful discussions
with M.~L\"uscher, G.C.~Rossi and R.~Sommer.

% \EPSFIGURE{filename.eps}                       % if you need
% {Text of the caption.\label{figlabel}}         % to put a figure

% \TABLE{\begin{tabular}...
%       ....
%       \end{tabular}%
%       \caption{Text of the caption             % If you need to put
%                of the table.\label{tablabel}}} % a table

\end{document}